\documentclass[aps,prl,reprint,superscriptaddress,showpacs,amsmath,amssymb,preprintnumbers,english]{revtex4-1}
\usepackage{graphicx}% Include figure files
\usepackage{dcolumn}% Align table columns on decimal point
\usepackage{bm}% bold math
\usepackage{babel}
\usepackage[babel=true]{csquotes}
\usepackage{enumerate}

\begin{document}

\preprint{Revised}

\title{Surface composition of BaTiO$_3$/SrTiO$_3$(001) films grown by atomic oxygen plasma assisted molecular beam epitaxy}



\author{A. Barbier}
%\thanks{}
\affiliation{CEA, IRAMIS, SPCSI, F-91191 Gif-sur-Yvette, France}
\author{C. Mocuta}
%\thanks{}
\affiliation{Synchrotron SOLEIL, L'Orme des Merisiers Saint-Aubin, BP 48, F-91192 Gif-sur-Yvette Cedex, France}
\author{D. Stanescu}
%\thanks{}
\affiliation{CEA, IRAMIS, SPCSI, F-91191 Gif-sur-Yvette, France}
\author{P. Jegou}
%\thanks{}
\affiliation{CEA, IRAMIS, SPCSI, F-91191 Gif-sur-Yvette, France}
\author{N. Jedrecy}
%\thanks{}
\affiliation{Institut des Nano Sciences de Paris, UPMC-Sorbonne Universit\'es, CNRS-UMR7588, 75005 Paris, France}
\author{H. Magnan}
%\thanks{}
\affiliation{CEA, IRAMIS, SPCSI, F-91191 Gif-sur-Yvette, France}


\date{\today}

\begin{abstract}
We have investigated the growth of BaTiO$_3$ thin films deposited on pure and 1\% Nb-doped SrTiO$_3$(001) single crystals using atomic oxygen assisted molecular beam epitaxy (AO-MBE) and dedicated Ba and Ti Knudsen cells. Thicknesses up to 30 nm were investigated for various layer compositions. We demonstrate 2D growth and epitaxial single crystalline BaTiO$_3$ layers up to 10 nm before additional 3D features appear; lattice parameter relaxation occurs during the first few nanometers and is completed at $\sim$10 nm. The presence of a Ba oxide rich top layer that probably favors 2D growth is evidenced for well crystallized layers. We show that the Ba oxide rich top layer can be removed by chemical etching. The present work stresses the importance of stoichiometry and surface composition of BaTiO$_3$ layers, especially in view of their integration in devices.
\end{abstract}

%\keywords{}

\maketitle

%\section{}
%\subsection{}
%\subsubsection{}

\section{Introduction}

 A renewed interest on ferroelectric materials has emerged recently for the development of nonvolatile random access memories and novel applications in the field of spintronics \cite{Fert:2008}. They are expected to provide a way to elaborate \enquote{on demand} multiferroic materials. In a multiferroic several ferroic orders coexist and may be coupled like \emph{e.g.} ferroelectric and ferro- (or antiferro-) magnetic long-range orders. This opens exciting prospects like the control of the magnetic state of an individual device by an electric field (and/or the control of the ferroelectric state by a magnetic field) \cite{Ramesh:2007}. The cross couplings between ferroelectric, ferromagnetic, ferroelastic and ferrotoroidic orders at the nanometer scale, called nanoferronics, open new routes for original applications \cite{Bibes:2012}.

 Within this context BaTiO$_3$ rapidly emerged as a material of choice since it stands for \enquote{the model ferroelectric} perovskite type material since decades. The bulk material has widespread applications in ferroelectric, piezoelectric, optoelectronic, photoferroelectric and photorefractive devices \cite{Moulson:1992,Pohanka:1988,Warren:1994,Klein:1988}. Moreover, the coupling between ferroelectricity and magnetism through interface bonding has been predicted theoretically for Fe/BaTiO$_3$ \cite{Duan:2006} and has been experimentally observed, although attributed to interface strain coupling \cite{Sahoo:2007}. The manipulation of the ferromagnetic magnetization via electric fields was also reported in Ni/BaTiO$_3$ hybrid structures where the application of an electric field to the ferroelectric BaTiO$_3$ induces elastic strain which is transferred into the ferromagnetic Ni layer, affecting its magnetization due to inverse magnetostriction \cite{Geprags:2010}. The inclusion of a single ferroelectric film of BaTiO$_3$ in a spin valve configuration showed genuine tunnel electro-resistance effects as the result of very low thickness and highly strained state of the layer \cite{Garcia:2009}. Recently, ferroelectricity and ferromagnetism coexistence was reported for Fe-doped BaTiO$_3$ \cite{Xu:2009}. This last system has also recently shown interesting photochemical activity making promise in the field of hydrogen production \cite{Upadhyay:2011}.

 Most novel applications including BaTiO$_3$ are now based on thin films of nanometer thickness. Chemistry based methods to produce BaTiO$_3$  thin films using nanopowders synthesized by the polymeric precursor route \cite{Upadhyay:2011} are efficient but not well suited for the thin films dedicated to nanoferronics applications. Within a surface science approach pulsed laser ablation is very popular to grow such films, the technique relies however on the composition of the target and the presence or not of additional oxygen during growth. This approach was used to estimate the role of the relative Ba:Ti non stoichiometries in Ba$_x$Ti$_{1-x}$O$_3$ layers \cite{Suzuki:2000} and showed that epitaxial growth is only possible close to the perovskite BaTiO$_3$ composition. Molecular beam epitaxy was also used in addition to molecular oxygen \cite{Niu:ME2011} and it could be shown that the ferroelectric properties strongly depend on the oxygen content in the layers.

 Considering thin film applications the question of surface/interfacial composition becomes crucial and deserves a detailed investigation which is the purpose of the present paper. The BaTiO$_3$ thin films were grown by atomic oxygen assisted molecular beam epitaxy (AO-MBE), an approach which enables an individual dosing of the metal ions involved and avoids the growth of oxygen deficient layers, that have been shown to exhibit reduced ferroelectrical properties \cite{Niu:ME2011}. After describing the experimental procedure we will present our results that will be discussed before concluding remarks will be drawn.

\section{Experimental}

The SrTiO$_3$(001) substrate samples were 10 mm $\times$ 10 mm or 5 mm $\times$ 5 mm squares of 1 mm thickness provided, aligned to $\pm$0.1$^{\circ}$, cut and polished by Crystal GmBH (Berlin, Germany). Pure SrTiO$_3$(001) and 1$\%$ Nb doped substrates were considered; the latter is known to have good electrical conductivity whereas pure crystals are highly insulating.  The BaTiO$_3$ growth was performed in a dedicated AO-MBE setup working in the 10$^{-10}$ mbar pressure range equipped with a RF plasma source, Auger electron spectroscopy (AES), reflection high energy electron diffraction (RHEED) and connected to a separate ultra-high vacuum (UHV) chamber with a X-ray photoemission spectrometry (XPS) apparatus. The cleanliness was checked by Auger Electron Spectroscopy at each step of the cleaning procedure and layer deposition. Optimal growth results required a first air annealing of the substrates in order to recrystallize the top surface atomic planes. The substrates were air annealed at 1250 K during 3 hours, the temperature was raised and decreased in 3 hours each. \emph{In situ} sample cleaning was achieved by $\sim$15 minutes outgassing at 600 K followed by $\sim$90 minutes exposure to a high brilliance oxygen plasma (power 350 W) with the sample held at 900 K - residual vacuum 5$\times$10$^{-8}$mbar (baratron 1.3 tr). These conditions proved to be efficient to fully remove the carbon contamination and to obtain sharp and well contrasted RHEED patterns typical for well ordered surfaces.

The Ba$_x$Ti$_{1-x}$O$_3$ layers were deposited by evaporation of metallic individual Ba and Ti from dedicated Knudsen cells assisted by a high brilliance (350 Watt - 10$^{-7}$ mbar oxygen - baratron 3.1 tr) atomic oxygen plasma. The growth rate was in the range 0.5 - 2 \AA /min. Tungsten and tantalum crucibles were used to evaporate Ti and Ba respectively. High purity 99.999$\%$ materials were used. Metal Ba was kept in oil before insertion in the chamber because of its high reactivity with water. The oil was removed using several successive cyclohexane (C$_6$H$_{12}$) ultrasound assisted baths; the still wet Ba metal was introduced in the MBE system just prior pumping. This procedure allows easy and fast outgassing of the Ba charge. Our approach has the great advantage of enabling independent Ti and Ba dosing, as well as oxygen dosing. For the present study, all samples were elaborated in the maximal possible oxidation conditions corresponding to the growth of hematite layers \cite{barbier2005,barbier2007} in the same MBE chamber, which are thermodynamically far above the required conditions for BaTiO$_3$. Any oxygen deficiency can thus be outruled in the present work.

\begin{table*}
\caption{\label{Samples}Summary of significative sample growth conditions and properties examined in the present work. The AES ratio is given by derivative peak to peak amplitude from Ti$_{LMM}$/Ba$_{MNN}$ and the XPS ratio (Ti$^{S}_{2p}$/Ba$^{S}_{3d}$) is derived from the integrated intensities ratio I(Ti$_{2p}$)/I(Ba$_{3d}$) corrected by the relevant Scofield coefficients. The AES and XPS spectra were recorded at 800 K and room temperature respectively. The XRD column reports the thicknesses as derived from specular reflectivity measurements. RHEED patterns are classified as follows : (G) good contrast, (R) rough (3D), (B) high background, (N) none (absent). Pure SrTiO$_3$(001) substrates were used for samples indexed B\# and 1$\%$ Nb doped substrates for the ones indexed NB\#. The indication +OH refers to an ethanol bath with ultrasounds for the sample described the line above.}
\begin{ruledtabular}
\begin{tabular}{cccccccc}
 Sample & T$_{Ti}$ & T$_{Ba}$ & Time & AES & XPS & XRD & RHEED \\
 { } & ($^o$C) & ($^o$C) & min. & Ti$_{LMM}$/Ba$_{MNN}$ & Ti$^{S}_{2p}$/Ba$^{S}_{3d}$ & (nm) &\\
 \hline
 BaTiO$_3$(001) & - & - & - & - & 1 & - & -\\
 B1 & 1265 & 367 & 60 & 0.70 & 0.32 & 4.5 & B\\
 B3 & 1264 & 388 & 75 & 1.21 & 0.54 & 5.46 & G\\
 B4 & 1294 & 394 & 75 & 1.67 & 0.69 & 8.27 & G+R\\
 B10 & 1327 & 390 & 75 & 2.8° & 1 & 6.67 & B+N\\
 B11 & 1289 & 390 & 180 & 0.95 & 0.44 & 9.37 & G\\
 B13 & 1251 & 385 & 180 & 1.29 & 0.74 & - & G\\
 B13+OH(25') & - & - & - & - & 1.1 & 9.1 & -\\
 B14 & 1269 & 387 & 180 & 1.2 & 0.74 & - & G\\
 B14+OH(10') & - & - & - & 2.04 & 0.97 & 9.8 & -\\
 NB1 & 1290 & 385 & 75 & 1.20 & 0.54 & 3.87 & G\\
 NB2 & 1314 & 408 & 92 & 1.4 & 0.75 & 8.5 & G\\
 NB11 & 1268 & 386 & 360 & 1.1 & - & 20 & G+R\\
\end{tabular}
\end{ruledtabular}
\end{table*}

During growth the samples were permanently kept at $\sim$900 - 950 K and rotated around the surface normal in order to ensure an homogeneous deposit. Lower temperatures were not able to favor epitaxial growth as expected from previous reports on the growth of BaTiO$_3$. The growth was monitored by RHEED in real time. The stoichiometry of the layers was checked by AES and XPS after the growth. The results reported here were derived from about 10 preliminary deposits on metal Cu for calibration purpose and 30 different SrTiO$_3$ substrates. We will discuss the results from the most significative samples that are listed with some of their growth parameters in table \ref{Samples}.

The synchrotron radiation X-ray measurements (reflectivity and wide angle X-ray scattering (WAXS)) were performed on DiffAbs beamline \cite{DIFFABS} at synchrotron-SOLEIL (Saint-Aubin, France). It is a bending magnet beamline supplying monochromatic photons in the E = 3.5 - 23 keV energy range ($\Delta$E/E $\sim$ 10$^{-4}$). The monochromaticity is ensured by a double crystal Si(111) monochromator. If the first Si crystal is flat, the second one is bendable in the sagittal direction, and thus allows a focusing of the X-ray beam in the horizontal direction at the sample position. Two mirrors placed on each side of the monochromator along the beam are used for harmonics rejection, ensuring a typical harmonics rejection rate of 10$^{-6}$ for the experiments. Both of them are bendable in longitudinal direction: the first one is tuning the incident X-ray beam divergence to match the acceptance of the Si(111) crystal of the monochromator, while the second one is focusing the X-ray beam in the vertical direction. The result is a clean X-ray spot at the sample position, of typical size 300 $\times$ 300 $\mu$m$^2$ (full width at half maximum) and an integrated photon flux up to few 10$^{12}$ ph./s at 8 keV photon energy. These reported values can be seen as typical for most of the experiments, with the photon flux decaying slightly (less than 1 decade for the lower energies) when the borders of the above mentioned energy range are approached. The sample was mounted on a goniometer head with motorized lateral translation tables (1 $\mu$m accuracy) and placed on the 6-circles diffractometer (Kappa geometry). This ensures all the necessary movements for sample placing, aligning into the X-ray beam and diffraction measurements (grazing or wide angle). All the angular movements have an accuracy of 0.002$^{\circ}$. A Cyberstar scintillation (LaCl$_3$) point detector was used for all X-ray scattering experiments, ensuring high counting rates with low counting loss ($\sim$ 3\% at 10$^{6}$ counts/s). Two pairs of slits are mounted in front of the detector and ensure (i) the angular resolution (detection slits) and (ii) probing only an area on the center of the sample (scatter slits). Typically the angular resolution used for the reflectivity and WAXS measurements is of 0.017$^{\circ}$ (3$\times$10$^{-4}$ rad). In the case of the pole figure measurements, this angular resolution is degraded by a factor of 10.

Complementary laboratory X-ray diffraction and reflectivity measurements were performed on a Panalytical X'Pert MRD apparatus equipped with Cu K$\alpha$ source. The AFM measurements were performed on a VEECO Dimension 3100 Atomic Force Microscope. The electric polarization measurements were performed with a aixACCT Easy Check 300 TF Analyzer.

\section{Results}

\begin{figure}
 \includegraphics[angle=0, width=8cm]{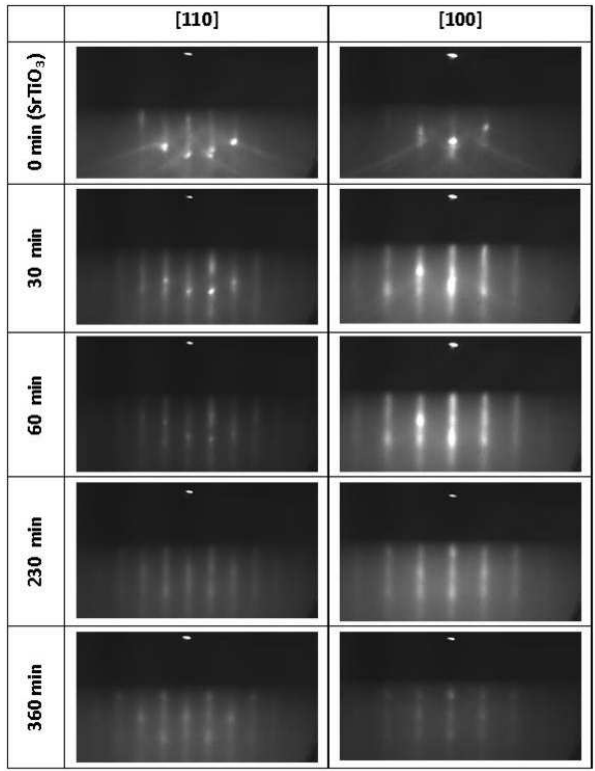}%
\caption{\label{RHEED}(Color online) (Top panel) Typical RHEED patterns, along [110] and [100] azimuths, observed during the growth of BaTiO$_3$ on SrTiO$_3$(001) for increasing thicknesses (sample NB11). The total thickness is $\sim$20nm. (Bottom panel) Evolution of the BaTiO$_3$/SrTiO$_3$(001) in-plane lattice parameter derived from RHEED patterns recorded during growth at a sample temperature of 900 K. Horizontal dotted and dashed lines indicate the SrTiO$_3$ and BaTiO$_3$ bulk cubic lattice parameters at 900 K.}
\end{figure}

For all layers the initial sharp RHEED SrTiO$_3$(001) pattern slightly degrades during the first 15 minutes of deposition, independently of further well contrasted 2D patterns or not. In optimal growth conditions we note that (i) very well contrasted RHEED patterns with intense 2D strikes are obtained (figure \ref{RHEED}) up to 10 nm thickness; (ii) above 10 nm faint 3D spots start to appear in the patterns indicating the onset of 3D crystallites on the top surface which can be more or less present from one sample to another and (iii) further deposition results in increasing 3D features along with still well defined strikes indicating a relatively flat average surface. Excess by 20\% Ti or Ba with respect to the 1:1 stoichiometry conditions are unfavorable to well contrasted RHEED patterns. Deposition with an excess of Ba results in high background (sample B1 - table \ref{Samples}). Excess Ti during growth is even more detrimental and leads rapidly to rough patterns (sample B4 - table \ref{Samples}) and finally to complete pattern vanishing (sample B10 - table \ref{Samples}) and high background.  The observation of well contrasted RHEED patterns remains however a modest criterion; changes in composition of at least $\pm$20$\%$ are needed to observe noticeable changes. However, a fairly more pronounced degradation is observed in the case of Ti rich growth conditions.

\begin{figure}
 \includegraphics[angle=0, width=8cm]{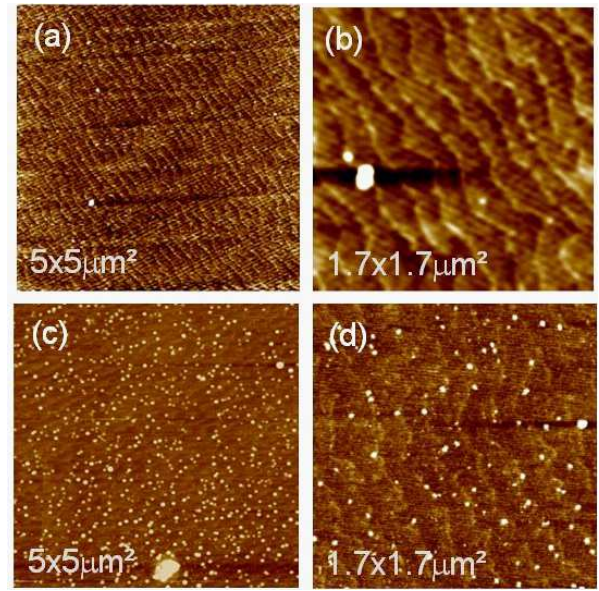}%
\caption{\label{AFM}(Color online) AFM images taken on BaTiO$_3$/SrTiO$_3$(001) deposits having well contrasted RHEED patterns during growth. (a) 5$\mu$m$\times$5$\mu$m image, \enquote{as grown} sample B1 (thickness 4.5 nm), (b) 1.7$\mu$m$\times$1.7$\mu$m image, \enquote{as grown} sample B1, (c) 5$\mu$m$\times$5$\mu$m image, \enquote{as grown} sample B14 (thickness $\sim$ 10 nm) and (d) 1.7$\mu$m$\times$1.7$\mu$m image, sample B14 after a 10 min. ethanol bath with ultrasound.}
\end{figure}

The morphological behavior qualitatively observed by RHEED was further investigated by direct space imaging. The transition from a step flow type 2D growth toward a 3D mode showing clusters at many places of the surface could be confirmed by \emph{ex situ} AFM (Atomic Force Microscopy) for images recorded on several places and of a few square microns in size, as shown in figure \ref{AFM}(a) and (b). The onset of cluster growth, on top of a still smooth layer, occurs typically around 10 nm thickness (see figure \ref{AFM}(c)).

For optimal growth conditions, the in-plane lattice parameter evolution has been investigated using integrated line profiles extracted from the RHEED patterns (figure \ref{RHEED}-bottom). This evolution has to be discussed in the light of the lattice parameters of SrTiO$_3$ and BaTiO$_3$ at the growth temperature of 900 K. The thermal expansion coefficient of strontium titanate has been investigated in detail \cite{Ligny:Phys.Rev.B.1995}. Because of its simple cubic crystalline structure its thermal expansion coefficient has been found almost constant up to 1800 K with a value of $\Delta$V/V$_{T=300K}$ = 3.23$\times$10$^{-5}$/K. To the contrary the linear thermal expansion coefficient of BaTiO$_3$ shows three anomalies, connected with the phase transitions Pm3m $\rightarrow$ P4mm $\rightarrow$ C2mm $\rightarrow$ R3m, at 401.8 K, 299.5 K and 216.1 K respectively \cite{Gorev:Phys.Rev.B.1995}. However, above the highest phase transition temperature the linear thermal expansion coefficient of BaTiO$_3$ can be approximated by its average value of $\Delta$L/L$_{T=300K}$ = 13$\times$10$^{-6}$/K \cite{Rao:BMS.1997,Schubert:Appl.Phys.Lett.2003,Siegert:MRS.2000}. At room temperature the lattice parameters of cubic SrTiO$_3$ and BaTiO$_3$ are 0.3905 nm and 0.3992 nm, we thus obtain 0.393 nm and 0.403 nm respectively at 900 K. From figure \ref{RHEED}-bottom it can be seen that the in-plane lattice parameter during growth remains pseudomorphic only within the first growing nanometer. Full lattice relaxation is completed at about 10 nm within the error bars of our measurement. This result is consistent with the TEM study by Sun et al. \cite{Sun:APL2004} that showed that the strain relaxation proceeds in BaTiO$_3$/SrTiO$_3$ via the introduction of an increasing density of dislocation lines above 2 nm thickness. Our results were confirmed by $\emph{ex situ}$ laboratory X-ray diffraction measurements on samples B1, B4, NB2 and B14 on asymmetrical reflections. Above 8 nm the lattice parameters are found fairly relaxed with a = (0.399 $\pm$ 0.001) nm and c = (0.4038 $\pm$ 0.0001) nm to be compared with the bulk BaTiO$_3$ values of a = 0.3992 nm and c = 0.4036 nm. The slower lattice parameter relaxation reported by Niu et al. \cite{Niu:ME2011} may found its explanation in the oxygen vacancies probably included in their films because of the use of molecular (and much less reactive) oxygen.

\begin{figure}
 \includegraphics[angle=0, width=8cm]{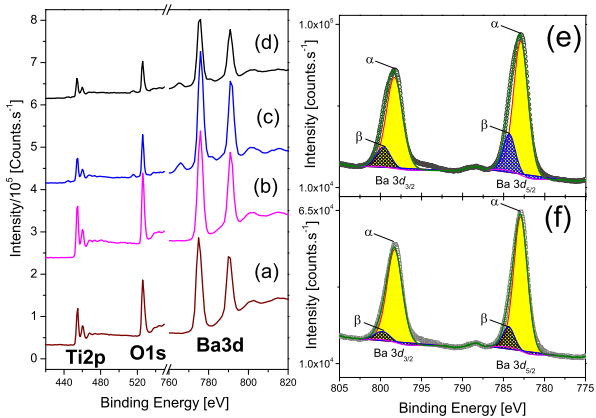}%
\caption{\label{XPS}(Color online)(Left Panel) Photoemission scans of (a) a BaTiO$_3$ single crystal Ti$^{S}_{2p}$/Ba$^{S}_{3d}$ = 1, (b) sample B10 (75 min deposit) exhibiting  Ti$^{S}_{2p}$/Ba$^{S}_{3d}$ = 1 but without any RHEED pattern, (c) as grown sample B14 (180 min deposit) with Ti$^{S}_{2p}$/Ba$^{S}_{3d}$ = 0.74 and sharp RHEED pattern and (d) sample B14 after a 10 min ethanol bath (with ultrasounds) Ti$^{S}_{2p}$/Ba$^{S}_{3d}$ = 0.97. (Right panel) Ba 3$\emph{d}_{5/2}$ and Ba 3$\emph{d}_{3/2}$ photoelectron peaks (sample B14, 180 min, $\sim$ 10 nm) measured with a pass energy of 20 eV for (e) as grown BaTiO$_3$ layer and (f) after a 10 min bath in ethanol with ultrasounds. Symbols stand for experimental data, filled areas correspond to $\alpha$ components and patterned areas to $\beta$ components, thin line for Shirley type background and thick straight line for the best fit. The best fit parameters are reported in table \ref{tableBa3d}.}
\end{figure}

XPS is a well established tool for chemical dosing at surfaces. It is a photon-in electron-out technique with moderate sensitivity to charge build-up contrarily to AES which is fully electron based and suffers heavily from charging effects of insulators. For the purpose of a reference comparison, a BaTiO$_3$(001) single crystal has been measured in our system. The c-oriented single crystal was provided, cut, polished and aligned to better than 0.1$^{\circ}$ by Crystal GmBH (Berlin, Germany). It was made from a (100) single crystal by electrical poling, the single domain fraction was not less than 80$\%$. The composition was estimated from the Ti$_{2p}$ and Ba$_{3d}$ photoemission lines. By convention, we note Ti$^{S}_{2p}$/Ba$^{S}_{3d}$ the integrated intensities ratio corrected by the corresponding Scofield factors for relative sensitivity (cross section). For the single crystal we obtain Ti$^{S}_{2p}$/Ba$^{S}_{3d}$=1 and thus the expected 1:1 ratio for Ti:Ba (see table \ref{Samples} and figure \ref{XPS}-(a)). This indicates that the difference in mean free paths between the Ba$_{3d}$ and Ti$_{2p}$ lines does not significatively impact on the corrected intensities for bulk BaTiO$_3$; this assumption should, by extension, also remain true for layers with thicknesses below the Ba$_{3d}$ and Ti$_{2p}$ mean free paths, which is the case of most of our samples.  Unfortunately the tetragonal BaTiO$_3$ single crystal changes the phase when its temperature is lower than 286 K or higher than 398 K. Thus AES reference measurements that would require elevated temperature to promote charge dissipation were not possible for the single crystal. For our thin films AES spectra measurements shown to be possible at temperatures above 700 K.

\begin{figure}
 \includegraphics[angle=0, width=8cm]{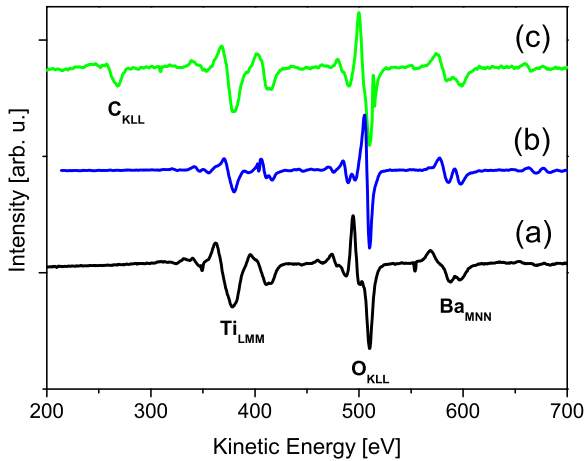}%
\caption{\label{AES}(Color online) Auger scans for several Ba$_x$Ti$_{1-x}$O$_3$ layers deposited on SrTiO$_3$(001). (a) Sample B10 with I(Ti$_{2p)}$/I(Ba$_{3d}$) photoemission line intensities ratio of 1 and Ti$_{LMM}$/Ba$_{MNN}$=2.8; (b) Sample B14 as grown with Ti$_{LMM}$/Ba$_{MNN}$ = 1.2 and (c) sample B14 after ethanol bath (with ultrasound) for 30 min and with Ti$_{LMM}$/Ba$_{MNN}$ = 2.}
\end{figure}

Sample B10 was grown in order to reproduce a Ti$^{S}_{2p}$/Ba$^{S}_{3d}$ photoemission lines ratio of 1 (see table \ref{Samples} and figure \ref{XPS}-(b)). Importantly, this sample revealed several major problems: (i) a rapid vanishing of the RHEED pattern during growth leaving only a strong background and (ii) an AES Ti$_{LMM}$/Ba$_{MNN}$ ratio of 2.8 (figure \ref{AES}-(a)), far from the value of 1.2 reported by Shimoyama et al. \cite{Shimoyama:2001} obtained for samples grown without external oxygen. To the contrary sample B14, that shows a XPS Ti$^{S}_{2p}$/Ba$^{S}_{3d}$ ratio of only 0.74, has the AES  Ti$_{LMM}$/Ba$_{MNN}$ ratio of 1.2 suggested in reference \onlinecite{Shimoyama:2001} as standing for a stoichiometric BaTiO$_3$ films. Moreover this last layer showed sharp, well contrasted and perovskite structured RHEED pattern up to 10 nm thickness. The deconvolution of the Ba \emph{3d} lines of sample B14 reveals the presence of two components corresponding to two chemical states labeled $\alpha$ and $\beta$ in figure \ref{XPS}-(e); the best fit parameters are reported in table \ref{tableBa3d}. The Ba 3$d$ line shapes of BaTiO$_3$ surfaces has been extensively considered by Mukhopadhyay and Chen \cite{Mukhopadhyay:1995} who investigated in detail the $\alpha$:$\beta$ ratio for several surface processing indicating that the $\beta$ component is related to the topmost surface region. After an ethanol bath the $\beta$ component decreases. The usefulness of this treatment will be discussed later in this paper.

\begin{table}
\caption{\label{tableBa3d}Best fit parameters for Ba 3d photo-emission lines deconvolution for a 10 nm thick \enquote{as grown} and ethanol cleaned  BaTiO$_3$/SrTiO$_3$ sample. E stands for binding energy, $\Delta$ for peak width (full width at half maximum values), $\rho$ for the relative weight of each line and $\Phi$ the total relative fraction of $\alpha$ and $\beta$ lines.}
\begin{ruledtabular}
\begin{tabular}{ccccccccccc}
 { } & \vline & \multicolumn{3}{c}{Ba 3d3/2}& \vline & \multicolumn{3}{c}{Ba 3d5/2} &  \vline & $\Phi$  \\
 Peak & \vline & E [eV] & $\Delta$ [eV] & $\rho$ [\%] & \vline & E [eV] & $\Delta$ [eV] & $\rho$ [\%] & \vline & \% \\
 \hline
  { } & \vline & \multicolumn{9}{c}{As grown}\\
 \hline
 $\alpha$ & \vline & 798.30 & 2.43 & 43.42 & \vline & 782.91 & 2.19 & 39.75 & \vline & 83.2\\
 $\beta$ & \vline & 799.60 & 2.07 & 8.04 & \vline & 784.31 & 1.76 & 8.79 & \vline & 16.8\\
 \hline
  { } & \vline & \multicolumn{9}{c}{+ 10 min Ethanol bath}\\
 \hline
 $\alpha$ & \vline & 798.27 & 2.30 & 47.75 & \vline & 782.90 & 2.05 & 42.70 & \vline & 90.5\\
 $\beta$ & \vline & 799.77 & 2.23 & 4.10 & \vline & 784.37 & 1.72 & 5.46 & \vline & 9.5\\

\end{tabular}
\end{ruledtabular}
\end{table}

A close investigation of the Auger spectra (figure \ref{AES}) shows that the shapes of Ti$_{LMM}$ and Ba$_{MNN}$ peaks correspond to fully oxidized ionic species and compare fairly well with TiO$_2$ \cite{Henrich:1978,Solomon:1975} and BaO \cite{Verhoeven:1979} reference spectra. This observation is valid for all \enquote{as grown} layers (measured \emph{in situ} just after growth at 700 K) within the range of stoichiometries investigated in the present work and also after exposure of the surface to air and/or to an ethanol bath (figure \ref{AES}). The later treatment lets appear additional carbon contamination only due to the exposure to air.

\begin{figure*}
 \includegraphics[angle=0, width=16cm]{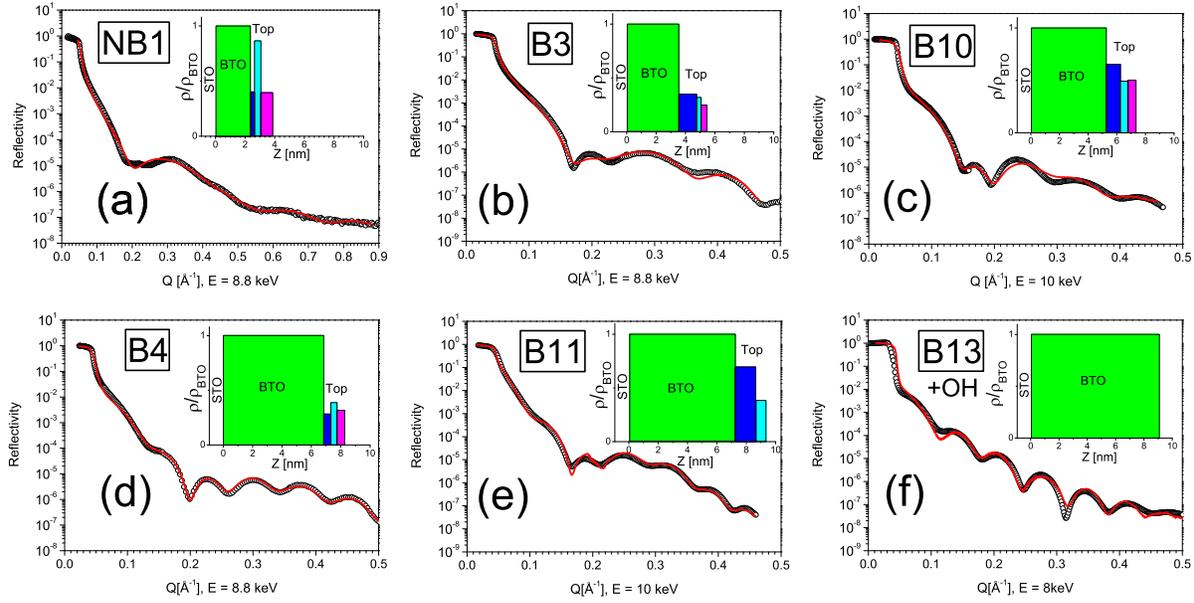}%
\caption{\label{Refl}(Color online) Reflectivity measurements ($\circ$) and best fit ($-$) for samples with increasing BaTiO$_3$ (BTO) thickness deposited on StTiO$_3$ (STO). (a) sample NB1, 3.87 nm; (b) B3, 5.46nm; (c) sample B10, 6.67 nm; (d) B4, 8.27 nm; (e) sample B11, 9.37; (f) sample B13, 9.1 nm after ethanol bath. For each situation an inset shows the density profile corresponding to the best fit.}
\end{figure*}

High resolution reflectivity curves were recorded up to large momentum transfer values (Q) for \enquote{as-grown} samples exhibiting variously contrasted RHEED patterns in the range 3 to 10 nm (figure \ref{Refl}(a-e)). To make the comparison easier all scans are represented with respect to Q coordinates expressed in {\AA}$^{-1}$ although several beam energies were used; the photon beam energy is recalled in all figures. Because X-ray thickness Kiessig fringes derive from a kinematical (2 waves) process it is possible to easily model the corresponding structure especially when data are taken with high signal to noise ratio over a large signal dynamical range as it is possible on synchrotron radiation beamlines. Reproducing the reflectivity curves \cite{Simul} using a single layer obviously failed for all thicknesses and all \enquote{as grown} investigated layers. In order to reproduce the experimental data, the samples were modeled by a core BaTiO$_3$ layer and a number of density \enquote{free} top layers. An additional density \enquote{free} layer was only introduced after failure to reproduce the experimental data with a lower number of \enquote{free} layers. All reflectivity curves could be reproduced using 2 or 3 top layers with their density left as a free parameter. The substrate roughness parameter was found negligible. The top layer roughnesses were in the range 0.4 to 0.6 nm r.m.s. and are not individually reported. Insets in figure \ref{Refl} sketch the density profile of the best fit model in each case. For all \enquote{as grown} samples a top layer with reduced density is necessary to reproduce the experimental data. The density variations between top layers remain in most cases very modest. Although their individual introduction highly improves the fit, their distribution is thus probably not very significative. More interestingly, the average density of the top layers lies in the 30 - 70\% range with respect to the BaTiO$_3$ density and appears as the general feature here. We thus estimate that the relevant parameter is the total layer thickness of reduced density.

\begin{figure}
 \includegraphics[angle=0, width=6cm]{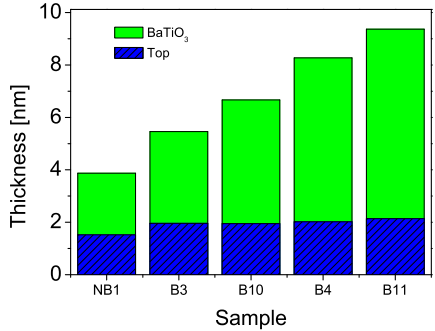}%
\caption{\label{Compos}(Color online) Layer composition as deduced from reflectivity measurements. For each sample considered in figure \ref{Refl} the total thickness is decomposed in BaTiO$_3$ and reduced density top surface layer thicknesses.}
\end{figure}

The relative ratio between the total thickness and the deficient top layer thickness is reported in figure \ref{Compos} for the \enquote{as grown} samples. The less dense top layer thickness (y) increases linearly with the total thickness (x) with a very small slope, following y =(1.3 $\pm$ 0.2) + (0.09 $\pm$ 0.03) $\cdot$ x for (4 nm $<$ x $<$ 10 nm). For the thinner layers, the proportion of the reduced density top layer with respect to the total thickness, is fairly larger.  Its predominance in the early stages of growth would be consistent with the systematic initial degradation of the RHEED patterns during the first $\sim$20 minutes before the contrast improves again.

\begin{figure}
 \includegraphics[angle=0, width=8cm]{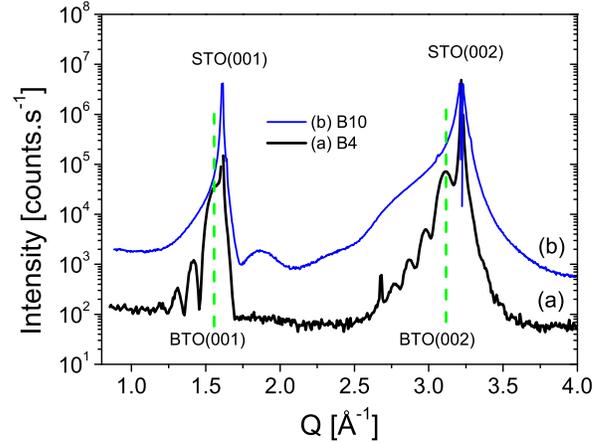}%
\caption{\label{WXRD}(Color online) Wide angle X-ray diffraction along specular direction for (a) sample B4 (8.3 nm with Ti$^{S}_{2p}$/Ba$^{S}_{3d}$ = 1.67 and Ti$^{S}_{2p}$/Ba$^{S}_{3d}$=0.69) and (b) sample B10 (6.7 nm with I(Ti$_{LMM}$)/I(Ba$_{MNN}$) = 2.8 and Ti$^{S}_{2p}$/Ba$^{S}_{3d}$=1).}
\end{figure}

Wide angle specular X-ray measurements confirmed the very poor crystalline quality of sample B10 (figure \ref{WXRD}) for which the BaTiO$_3$ 002 Bragg peak appears to the best in the form of a bump on the low Q side of the SrTiO$_3$ 002 Bragg peak. Comparatively samples containing less Ti, like sample B4 in figure \ref{WXRD},  show well ordered structures, well defined Bragg peaks and out-of-plane mosaic spread as small as 0.03 degrees. They also exhibit well defined wide angle oscillations that correspond to the coherent thickness of the layer \emph{i.e.} the thickness having a defined registry with the substrate. Interestingly, the coherent thickness was found systematically about 2 nm smaller than the thickness deduced from the reflectivity Kiessig thickness oscillations which is consistent with the reflectivity analysis.

\begin{figure}
 \includegraphics[angle=0, width=8cm]{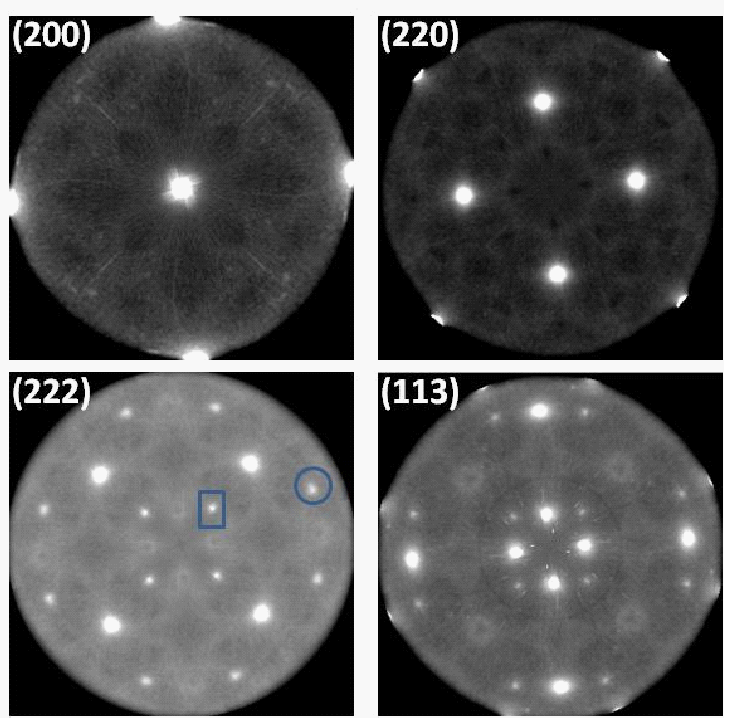}%
\caption{\label{Poles}(Color online) 200, 220, 222 and 113 pole figures measured on sample B13. The angular range was 0 $<$ $\phi$ $<$ 120$^{\circ}$ and 0 $<$ $\psi$ $<$ 89$^{\circ}$ ($\psi$=0 corresponds to the map center) for the azimuth and polar angles respectively. The figures were symmetrized over 90$^{\circ}$ after checking for one map an identical result for a measurement performed over a range 0 $<$ $\phi$ $<$ 180$^{\circ}$.  The top surface layer has been removed using a 25 minutes ethanol bath with ultrasounds. The positions indicated in pole figure (222) by a square and a circle corresponds to the SrTiO$_3$ substrate Bragg peaks 113 and 311 respectively; this can be easily understood by considering the symmetry of pole figure (113) in which the 222 substrate Bragg features also appear.}
\end{figure}

The structural characterization of our layers was completed by pole figure measurements. Within a pole measurement experiment the scattering angle is fixed to correspond to a given family of reticular planes whereas azimuthal ($\phi$) and polar ($\psi$) angles are varied over all possible positions giving thus the opportunity for all planes, of the given reticular family, to diffract. The patterns easily reveal the presence of textures or misoriented crystallites included in the film. Figure \ref{Poles} shows the pole figures recorded for the 200, 220, 222 and 113 Bragg peaks of sample B13 which exhibited well contrasted RHEED pattern during growth along with an AES Ti$_{LMM}$/Ba$_{MNN}$ ratio of 1.29 and an XPS ratio Ti$^{S}_{2p}$/Ba$^{S}_{3d}$ of 0.74 after growth. The pole figures were recorded after etching of the sample with ethanol. The pole figures include exclusively reflections corresponding to perfectly epitaxial BaTiO$_3$(001) without any detectable texture or misorientation. The weaker spots in the pattern were all identified as traces belonging to the substrate scattering. The grown layer, within these growth parameters, can thus be considered as single crystalline and epitaxial.

\begin{figure}
 \includegraphics[angle=0, width=8cm]{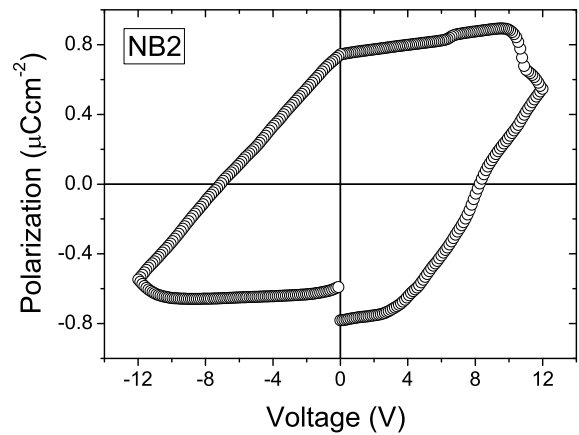}%
\caption{\label{Pol}(Color online) Electric polarization of sample NB2. An additional Au/Cr contact was evaporated on top of the sample in order to make the measurement possible.}
\end{figure}

The ferroelectric properties of BaTiO$_3$ are linked to the existence of a well-defined perovskite tetragonal structure, where Ti and O ions are displaced with respect to Ba cations so that the positive and negative ions barycenters do not coincide. The onset of this physical property is an important criterion to validate the quality and structure of the grown thin films. The ferroelectric behavior of layers having well contrasted RHEED patterns during growth was tested using sample NB2. Several 100/10 nm thick Au/Cr disk shaped top electrodes with a diameter ranging from 70 to 750 $\mu$m were evaporated on the surface, the Nb-doped SrTiO$_3$ substrate playing bottom electrode role. A typical hysteresis loop is provided in figure \ref{Pol}. It looks similar to that usually obtained for ultra thin films ($\sim$10 nm), where leakage currents hamper the determination of the ferroelectric contribution \cite{Nagarajan:2004,Allibe:2012}. Moreover in our case, the measurements are made difficult by the fact that bias is applied by a direct contact to the substrate, a situation that probably generates a non uniform electrical field.

\section{Discussion}
\label{Discussion}

Reproducing an \emph{in situ} Ti$^{S}_{2p}$/Ba$^{S}_{3d}$ photoemission lines ratio of 1 for thin BaTiO$_3$ films could naively appear as the relevant strategy in order to produce stoichiometric BaTiO$_3$ thin films especially since an abundant number of paper in literature claim that XPS was the method confirming the thin film stoichiometry. Using independent Ti and Ba dosing this could be easily achieved with sample B10 but resulted in very poor crystalline structure and much a too large AES Ti$_{LMM}$/Ba$_{MNN}$ ratio with respect to reference \onlinecite{Shimoyama:2001}. A previous investigation of the effect of non stoichiometries of BaTiO$_3$ films elaborated by pulsed laser deposition was reported by Suzuki et al. \cite{Suzuki:2000} using targets of pre-defined stoichiometries. These authors showed that fair deviations from the ideal stoichiometry leads to microstructural changes that are highly detrimental to the long range order in the films. Both Ti and Ba excesses were shown to favor the growth of amorphous phases including nanocrystalline epitaxial grains. Hence, although sample B10 exhibited the expected Ti$^{S}_{2p}$/Ba$^{S}_{3d}$  ratio its amorphous structural behavior appeared incompatible with the presence of the sought epitaxial BaTiO$_3$ films. Intriguingly, samples B13 and B14 with adequate AES Ti$_{LMM}$/Ba$_{MNN}$ ratio (figure \ref{AES}-(b)) (with respect to Shimoyama's \cite{Shimoyama:2001} criterion) and well contrasted RHEED patterns show a XPS Ti$^{S}_{2p}$/Ba$^{S}_{3d}$ ratio of only 0.74 (figure \ref{XPS}). One may note that evaluating the actual AES ratio from reference data \cite{AES_Handbook} remains speculative in the present case because the inclusion of the cations in the oxide compound severely modifies the Ti \cite{Henrich:1978,Solomon:1975} and Ba \cite{Verhoeven:1979} Auger line shapes which comes in addition to the fact that the quantification of AES lines, depending on the homogeneity and morphology of the layers, is usually a complex problems by itself. For all \enquote{as grown} layers, whatever the RHEED pattern contrast, reflectivity measurements as well as WAXS coherent thicknesses indicate the presence of a top layer of different nature. It is less dense than BaTiO$_3$ and is incoherent with respect to the substrate. Combining these observations with the fact that the XPS Ti$^{S}_{2p}$/Ba$^{S}_{3d}$ ratio, for layers with well contrasted RHEED patterns, corresponds to an excess of Ba with respect to the single crystal reference data, we can reasonably propose that the surface region is Ba rich for these samples.

Following the assumption of a Ba rich surface oxide layer and knowing that barium oxide (BaO) is soluble in methanol and ethanol forming alkoxides, we attempted a simple ethanol bath assisted by ultrasound in order to remove the eventual barium oxide rich surface layer. This simple procedure remarkably modified the surface composition of our samples and reconciled all approaches for samples having well contrasted RHEED patterns during growth (sample B13 and B14). In particular :
\begin{enumerate}[(i)]
  \item Reflectivity and pole measurements on an etched sample showed a layer made of a single phase (figures \ref{Refl}(f) and \ref{Poles}).
  \item We observe a consistent increase of the AES Ti$_{LMM}$/Ba$_{MNN}$ ratio up to 2 (figure \ref{AES}(c)).
  \item The XPS Ti$^{S}_{2p}$/Ba$^{S}_{3d}$ ratio for sample B14 increases up to 1 (figure \ref{XPS}((d)) in par with the single crystal case.
  \item The surface $\beta$ component in the Ba$_{3\emph{d}}$ lines fairly decreases confirming the removal of surface contributions (figure \ref{XPS}).
  \item The size of the 3D clusters on top of the \enquote{as grown} surface decreases significatively (figure \ref{AFM}(d))
\end{enumerate}
We can thus conclude that the surface layer present for the \enquote{as grown} layers is a BaO rich layer which can efficiently be removed by ethanol etching. Moreover, the AFM images taken before and after ethanol baths, figures \ref{AFM}(c) and (d) respectively, indicate that BaO is also involved in the 3D clusters for the thicker layers.
This Ba rich top layer is present in all \enquote{as grown} samples. Since the XPS mean free path is larger or comparable to the total thickness, the XPS spectra quantification provides the mean composition of the complete layer, within the assumption of an homogeneous lateral composition. Therefore the underlying layer has necessarily a larger Ti:Ba ratio as compared to the surface region. For samples exhibiting well contrasted 2D RHEED patterns we show, through removal of the surface layer, that the underlayer has the Ti:Ba 1:1 ratio, equivalent to the single crystal. Hence, we understand that for sample B10, with a total chemical stoichiometry Ti:Ba of 1:1, the presence of a Ba rich surface layer imposes the presence of a fairly Ti deficient underlayer and in turn the loss of the RHEED pattern. The latter example stresses the fact that the Ba rich surface layer formation is favored during evaporation. The driving force could be Ba segregation or oxygen chemical potential. Let us note that after ethanol etching the samples were found stable over weeks without any further segregation at room temperature. This surface layer was not reported or not understood in previous reports, but several points indicate that it could have been present, like the often reported increase of the $\beta$ components in Ba$_{3\emph{d}}$ XPS spectra. It is not unlikely that the samples reported in reference \onlinecite{Shimoyama:2001} had a composition close to our samples. Consistently with our results, Li \emph{et al.} \cite{Li:APL2008} found a reduced surface density on thick BaTiO$_3$ films grown by laser molecular beam epitaxy (using molecular oxygen) as well as surface specific Ba 3$d$ core level shifts, but they concluded on the presence of surface relaxation supposed to originate from oxygen vacancies. From our present results we can suggest that the effect might be intrinsic to the growth of BaTiO$_3$ films and that it is probably not linked to oxygen vacancies alone. Interestingly, the Ba rich top layer exhibits the same crystallographic structure than the underlayer since equivalent lattice parameters are deduced from the RHEED patterns and the \emph{ex situ} X-ray measurements (which are sensitive to the whole film).

In agreement with previous studies the presence of well defined and contrasted RHEED patterns is a major criterion when growing BaTiO$_3$(001) thin films on SrTiO$_3$(001). However care has to be taken when using spectroscopies to assess the actual composition. Using AO-MBE a film with a stoichiometric BaTiO$_3$ core exhibits, after growth, an AES Ti$_{LMM}$/Ba$_{MNN}$ ratio of $\sim$1.2 and an XPS Ti$^{S}_{2p}$/Ba$^{S}_{3d}$ ratio of $\sim$0.7 which become $\sim$2 and $\sim$1 respectively after exposure to ethanol that dissolves the BaO rich top layer. Finally, stoichiometric BaTiO$_3$(001)/SrTiO$_3$(001) layers are more easily obtained in Ba rich conditions; it is likely that the BaO top layer acts as a surfactant layer during the growth since Ti rich conditions rapidly favor cluster growth. Large excess ($>$$\pm$20\%) in Ba or in Ti leads to amorphous phases.

\section{Conclusion}

We have investigated fully oxidized epitaxial BaTiO$_3$ thin films grown by atomic oxygen assisted molecular beam epitaxy at a substrate temperature of 900 K on SrTiO$_3$(001) using dedicated Ba and Ti Knudsen cells. Nonstoichiometries above $\pm$20\% were found severely detrimental to the crystalline long range order as attested by \emph{in situ} RHEED. In optimal growth conditions core BaTiO$_3$ epitaxial 2D films could be grown up to 10 nm thickness prior 3D crystallites appear with however excellent crystalline long range order. The lattice parameter is found completely relaxed for a 10 nm film thickness, which is a smaller value compared to studies using oxygen or cation deficiencies. The presence of a less dense Ba rich top layer has been identified by electron spectroscopies and X-ray reflectivity  with a thickness increasing slightly with respect to the total thickness. The possible presence of this top layer should be taken into account when multilayered devices are build and interface effects become important. Finally, we show that the top layer can easily be removed by a simple ethanol etching in ultrasound bath.

\section{Acknowledgments}

We acknowledge SOLEIL for provision of synchrotron radiation facilities and are grateful to the SOLEIL-DIFFABS beamline staff members whose efficient efforts have made these experiments possible. This work has been supported by the Region Ile-de-France in the framework of C'Nano IdF, the nanoscience competence center of Paris Region under grant MAEBA. N.J. wishes to thank L. Becerra (INSP) for the realization of Au electrodes, E. Lacaze (INSP)for the use of the AFM set-up, and S. Hidki (INSP)for the X-ray measurements.


\begin{thebibliography}{34}%
\makeatletter
\providecommand \@ifxundefined [1]{%
 \@ifx{#1\undefined}
}%
\providecommand \@ifnum [1]{%
 \ifnum #1\expandafter \@firstoftwo
 \else \expandafter \@secondoftwo
 \fi
}%
\providecommand \@ifx [1]{%
 \ifx #1\expandafter \@firstoftwo
 \else \expandafter \@secondoftwo
 \fi
}%
\providecommand \natexlab [1]{#1}%
\providecommand \enquote  [1]{``#1''}%
\providecommand \bibnamefont  [1]{#1}%
\providecommand \bibfnamefont [1]{#1}%
\providecommand \citenamefont [1]{#1}%
\providecommand \href@noop [0]{\@secondoftwo}%
\providecommand \href [0]{\begingroup \@sanitize@url \@href}%
\providecommand \@href[1]{\@@startlink{#1}\@@href}%
\providecommand \@@href[1]{\endgroup#1\@@endlink}%
\providecommand \@sanitize@url [0]{\catcode `\\12\catcode `\$12\catcode
  `\&12\catcode `\#12\catcode `\^12\catcode `\_12\catcode `\%12\relax}%
\providecommand \@@startlink[1]{}%
\providecommand \@@endlink[0]{}%
\providecommand \url  [0]{\begingroup\@sanitize@url \@url }%
\providecommand \@url [1]{\endgroup\@href {#1}{\urlprefix }}%
\providecommand \urlprefix  [0]{URL }%
\providecommand \Eprint [0]{\href }%
\@ifxundefined \urlstyle {%
  \providecommand \doi  [0]{\begingroup \@sanitize@url \@doi}%
  \providecommand \@doi [1]{\endgroup \@@startlink {\doibase
  #1}doi:\discretionary {}{}{}#1\@@endlink }%
}{%
  \providecommand \doi  [0]{doi:\discretionary{}{}{}\begingroup
  \urlstyle{rm}\Url }%
}%
\providecommand \doibase [0]{http://dx.doi.org/}%
\providecommand \Doi [0]{\begingroup \@sanitize@url \@Doi }%
\providecommand \@Doi  [1]{\endgroup\@@startlink{\doibase#1}\@@Doi}%
\providecommand \@@Doi [1]{#1\@@endlink}%
\providecommand \selectlanguage [0]{\@gobble}%
\providecommand \bibinfo  [0]{\@secondoftwo}%
\providecommand \bibfield  [0]{\@secondoftwo}%
\providecommand \translation [1]{[#1]}%
\providecommand \BibitemOpen [0]{}%
\providecommand \bibitemStop [0]{}%
\providecommand \bibitemNoStop [0]{.\EOS\space}%
\providecommand \EOS [0]{\spacefactor3000\relax}%
\providecommand \BibitemShut  [1]{\csname bibitem#1\endcsname}%
%</preamble>
\bibitem [{\citenamefont {Fert}(2008)}]{Fert:2008}%
  \BibitemOpen
  \bibfield  {author} {\bibinfo {author} {\bibfnamefont {A.}~\bibnamefont
  {Fert}},\ }\href@noop {} {\bibfield  {journal} {\bibinfo  {journal} {Angew.
  Chem. Int. Ed. (Nobel Lecture)},\ }\textbf {\bibinfo {volume} {45}},\
  \bibinfo {pages} {5956} (\bibinfo {year} {2008})}\BibitemShut {NoStop}%
\bibitem [{\citenamefont {Ramesh}\ and\ \citenamefont
  {Spaldin}(2007)}]{Ramesh:2007}%
  \BibitemOpen
  \bibfield  {author} {\bibinfo {author} {\bibfnamefont {R.}~\bibnamefont
  {Ramesh}}\ and\ \bibinfo {author} {\bibfnamefont {N.~A.}\ \bibnamefont
  {Spaldin}},\ }\href@noop {} {\bibfield  {journal} {\bibinfo  {journal}
  {Nature},\ }\textbf {\bibinfo {volume} {6}},\ \bibinfo {pages} {21} (\bibinfo
  {year} {2007})}\BibitemShut {NoStop}%
\bibitem [{\citenamefont {Bibes}(2012)}]{Bibes:2012}%
  \BibitemOpen
  \bibfield  {author} {\bibinfo {author} {\bibfnamefont {M.}~\bibnamefont
  {Bibes}},\ }\href@noop {} {\bibfield  {journal} {\bibinfo  {journal} {Nat.
  Mat.},\ }\textbf {\bibinfo {volume} {11}},\ \bibinfo {pages} {354} (\bibinfo
  {year} {2012})}\BibitemShut {NoStop}%
\bibitem [{\citenamefont {Moulson}\ and\ \citenamefont
  {Herbert}(1992)}]{Moulson:1992}%
  \BibitemOpen
  \bibfield  {author} {\bibinfo {author} {\bibfnamefont {A.~J.}\ \bibnamefont
  {Moulson}}\ and\ \bibinfo {author} {\bibfnamefont {J.~M.}\ \bibnamefont
  {Herbert}},\ }\href@noop {} {\emph {\bibinfo {title}
  {Electroceramics-Materials, Properties and Applications}}}\ (\bibinfo
  {publisher} {Chapman and Hall Publishers, London},\ \bibinfo {year}
  {1992})\BibitemShut {NoStop}%
\bibitem [{\citenamefont {Pohanka}\ and\ \citenamefont
  {Smith}(1988)}]{Pohanka:1988}%
  \BibitemOpen
  \bibfield  {author} {\bibinfo {author} {\bibfnamefont {R.~C.}\ \bibnamefont
  {Pohanka}}\ and\ \bibinfo {author} {\bibfnamefont {P.~S.}\ \bibnamefont
  {Smith}},\ }\href@noop {} {\emph {\bibinfo {title} {Electronic Ceramics}}},\
  edited by\ \bibinfo {editor} {\bibfnamefont {L.~M.}\ \bibnamefont
  {Levinson}}\ (\bibinfo  {publisher} {Marcel Dekker Inc. New York},\ \bibinfo
  {year} {1988})\BibitemShut {NoStop}%
\bibitem [{\citenamefont {Warren}\ and\ \citenamefont
  {Dimis}(1994)}]{Warren:1994}%
  \BibitemOpen
  \bibfield  {author} {\bibinfo {author} {\bibfnamefont {W.~L.}\ \bibnamefont
  {Warren}}\ and\ \bibinfo {author} {\bibfnamefont {D.}~\bibnamefont {Dimis}},\
  }\href@noop {} {\bibfield  {journal} {\bibinfo  {journal} {Appl. Phys.
  Lett.},\ }\textbf {\bibinfo {volume} {64}},\ \bibinfo {pages} {866} (\bibinfo
  {year} {1994})}\BibitemShut {NoStop}%
\bibitem [{\citenamefont {Klein}(1988)}]{Klein:1988}%
  \BibitemOpen
  \bibfield  {author} {\bibinfo {author} {\bibfnamefont {M.~B.}\ \bibnamefont
  {Klein}},\ }\href@noop {} {\emph {\bibinfo {title} {Photorefractive Materials
  and Their Applications}}},\ edited by\ \bibinfo {editor} {\bibfnamefont
  {P.}~\bibnamefont {Gunter}}\ and\ \bibinfo {editor} {\bibfnamefont {J.~P.}\
  \bibnamefont {Huignard}},\ Vol.~\bibinfo {volume} {61}\ (\bibinfo
  {publisher} {Springer-Verlag, Berlin},\ \bibinfo {year} {1988})\BibitemShut
  {NoStop}%
\bibitem [{\citenamefont {Duan}\ \emph {et~al.}(2006)\citenamefont {Duan},
  \citenamefont {Jaswal},\ and\ \citenamefont {Tsymbal}}]{Duan:2006}%
  \BibitemOpen
  \bibfield  {author} {\bibinfo {author} {\bibfnamefont {C.-G.}\ \bibnamefont
  {Duan}}, \bibinfo {author} {\bibfnamefont {S.}~\bibnamefont {Jaswal}}, \ and\
  \bibinfo {author} {\bibfnamefont {E.}~\bibnamefont {Tsymbal}},\ }\href@noop
  {} {\bibfield  {journal} {\bibinfo  {journal} {Phys. Rev. Lett.},\ }\textbf
  {\bibinfo {volume} {97}},\ \bibinfo {pages} {047201} (\bibinfo {year}
  {2006})}\BibitemShut {NoStop}%
\bibitem [{\citenamefont {Sahoo}\ \emph {et~al.}(2007)\citenamefont {Sahoo},
  \citenamefont {Polisetty}, \citenamefont {Duan}, \citenamefont {Jaswal},
  \citenamefont {Tsymbal},\ and\ \citenamefont {Binek}}]{Sahoo:2007}%
  \BibitemOpen
  \bibfield  {author} {\bibinfo {author} {\bibfnamefont {S.}~\bibnamefont
  {Sahoo}}, \bibinfo {author} {\bibfnamefont {S.}~\bibnamefont {Polisetty}},
  \bibinfo {author} {\bibfnamefont {C.~G.}\ \bibnamefont {Duan}}, \bibinfo
  {author} {\bibfnamefont {S.~S.}\ \bibnamefont {Jaswal}}, \bibinfo {author}
  {\bibfnamefont {E.~Y.}\ \bibnamefont {Tsymbal}}, \ and\ \bibinfo {author}
  {\bibfnamefont {C.}~\bibnamefont {Binek}},\ }\href@noop {} {\bibfield
  {journal} {\bibinfo  {journal} {Phys. Rev. B},\ }\textbf {\bibinfo {volume}
  {76}},\ \bibinfo {pages} {092108} (\bibinfo {year} {2007})}\BibitemShut
  {NoStop}%
\bibitem [{\citenamefont {Gepr\"{a}gs}\ \emph {et~al.}(2010)\citenamefont
  {Gepr\"{a}gs}, \citenamefont {Brandlmaier}, \citenamefont {Opel},
  \citenamefont {Gross},\ and\ \citenamefont {Goennenwein}}]{Geprags:2010}%
  \BibitemOpen
  \bibfield  {author} {\bibinfo {author} {\bibfnamefont {S.}~\bibnamefont
  {Gepr\"{a}gs}}, \bibinfo {author} {\bibfnamefont {A.}~\bibnamefont
  {Brandlmaier}}, \bibinfo {author} {\bibfnamefont {M.}~\bibnamefont {Opel}},
  \bibinfo {author} {\bibfnamefont {R.}~\bibnamefont {Gross}}, \ and\ \bibinfo
  {author} {\bibfnamefont {S.~T.~B.}\ \bibnamefont {Goennenwein}},\ }\href@noop
  {} {\bibfield  {journal} {\bibinfo  {journal} {Appl. Phys. Lett.},\ }\textbf
  {\bibinfo {volume} {96}},\ \bibinfo {pages} {142509} (\bibinfo {year}
  {2010})}\BibitemShut {NoStop}%
\bibitem [{\citenamefont {Garcia}\ \emph {et~al.}(2009)\citenamefont {Garcia},
  \citenamefont {Fusil}, \citenamefont {Bouzehouane}, \citenamefont
  {Enouz-Vedrenne}, \citenamefont {Mathur}, \citenamefont {Barthelemy},\ and\
  \citenamefont {Bibes}}]{Garcia:2009}%
  \BibitemOpen
  \bibfield  {author} {\bibinfo {author} {\bibfnamefont {V.}~\bibnamefont
  {Garcia}}, \bibinfo {author} {\bibfnamefont {S.}~\bibnamefont {Fusil}},
  \bibinfo {author} {\bibfnamefont {K.}~\bibnamefont {Bouzehouane}}, \bibinfo
  {author} {\bibfnamefont {S.}~\bibnamefont {Enouz-Vedrenne}}, \bibinfo
  {author} {\bibfnamefont {N.~D.}\ \bibnamefont {Mathur}}, \bibinfo {author}
  {\bibfnamefont {A.}~\bibnamefont {Barthelemy}}, \ and\ \bibinfo {author}
  {\bibfnamefont {M.}~\bibnamefont {Bibes}},\ }\href@noop {} {\bibfield
  {journal} {\bibinfo  {journal} {Nature},\ }\textbf {\bibinfo {volume}
  {460}},\ \bibinfo {pages} {81} (\bibinfo {year} {2009})}\BibitemShut
  {NoStop}%
\bibitem [{\citenamefont {Xu}\ \emph {et~al.}(2009)\citenamefont {Xu},
  \citenamefont {Yin}, \citenamefont {Lin}, \citenamefont {Xia}, \citenamefont
  {Wan}, \citenamefont {Yin}, \citenamefont {Bai}, \citenamefont {Du},\ and\
  \citenamefont {Liu}}]{Xu:2009}%
  \BibitemOpen
  \bibfield  {author} {\bibinfo {author} {\bibfnamefont {B.}~\bibnamefont
  {Xu}}, \bibinfo {author} {\bibfnamefont {K.~B.}\ \bibnamefont {Yin}},
  \bibinfo {author} {\bibfnamefont {J.}~\bibnamefont {Lin}}, \bibinfo {author}
  {\bibfnamefont {Y.~D.}\ \bibnamefont {Xia}}, \bibinfo {author} {\bibfnamefont
  {X.~G.}\ \bibnamefont {Wan}}, \bibinfo {author} {\bibfnamefont
  {J.}~\bibnamefont {Yin}}, \bibinfo {author} {\bibfnamefont {X.~J.}\
  \bibnamefont {Bai}}, \bibinfo {author} {\bibfnamefont {J.}~\bibnamefont
  {Du}}, \ and\ \bibinfo {author} {\bibfnamefont {Z.~G.}\ \bibnamefont {Liu}},\
  }\href@noop {} {\bibfield  {journal} {\bibinfo  {journal} {Phys. Rev. B},\
  }\textbf {\bibinfo {volume} {79}},\ \bibinfo {pages} {134109} (\bibinfo
  {year} {2009})}\BibitemShut {NoStop}%
\bibitem [{\citenamefont {Upadhyay}\ \emph {et~al.}(2011)\citenamefont
  {Upadhyay}, \citenamefont {Shrivastava}, \citenamefont {Solanki},
  \citenamefont {Choudhary}, \citenamefont {Sharma}, \citenamefont {Kumar},
  \citenamefont {Singh}, \citenamefont {Satsangi}, \citenamefont {Shrivastav},
  \citenamefont {Waghmare},\ and\ \citenamefont {Dass}}]{Upadhyay:2011}%
  \BibitemOpen
  \bibfield  {author} {\bibinfo {author} {\bibfnamefont {S.}~\bibnamefont
  {Upadhyay}}, \bibinfo {author} {\bibfnamefont {J.}~\bibnamefont
  {Shrivastava}}, \bibinfo {author} {\bibfnamefont {A.}~\bibnamefont
  {Solanki}}, \bibinfo {author} {\bibfnamefont {S.}~\bibnamefont {Choudhary}},
  \bibinfo {author} {\bibfnamefont {V.}~\bibnamefont {Sharma}}, \bibinfo
  {author} {\bibfnamefont {P.}~\bibnamefont {Kumar}}, \bibinfo {author}
  {\bibfnamefont {N.}~\bibnamefont {Singh}}, \bibinfo {author} {\bibfnamefont
  {V.~R.}\ \bibnamefont {Satsangi}}, \bibinfo {author} {\bibfnamefont
  {R.}~\bibnamefont {Shrivastav}}, \bibinfo {author} {\bibfnamefont {U.~V.}\
  \bibnamefont {Waghmare}}, \ and\ \bibinfo {author} {\bibfnamefont
  {S.}~\bibnamefont {Dass}},\ }\href@noop {} {\bibfield  {journal} {\bibinfo
  {journal} {J. Phys. Chem. C},\ }\textbf {\bibinfo {volume} {115}},\ \bibinfo
  {pages} {24373} (\bibinfo {year} {2011})}\BibitemShut {NoStop}%
\bibitem [{\citenamefont {Suzuki}\ \emph {et~al.}(2000)\citenamefont {Suzuki},
  \citenamefont {Nishi},\ and\ \citenamefont {Fujimoto}}]{Suzuki:2000}%
  \BibitemOpen
  \bibfield  {author} {\bibinfo {author} {\bibfnamefont {T.}~\bibnamefont
  {Suzuki}}, \bibinfo {author} {\bibfnamefont {Y.}~\bibnamefont {Nishi}}, \
  and\ \bibinfo {author} {\bibfnamefont {M.}~\bibnamefont {Fujimoto}},\
  }\href@noop {} {\bibfield  {journal} {\bibinfo  {journal} {Jpn. J. Appl.
  Phys.},\ }\textbf {\bibinfo {volume} {39}},\ \bibinfo {pages} {5970}
  (\bibinfo {year} {2000})}\BibitemShut {NoStop}%
\bibitem [{\citenamefont {Niu}\ \emph {et~al.}(2011)\citenamefont {Niu},
  \citenamefont {Yin}, \citenamefont {Saint-Girons}, \citenamefont {Gautier},
  \citenamefont {Lecoeur}, \citenamefont {Pillard}, \citenamefont {Hollinger},\
  and\ \citenamefont {Vilquin}}]{Niu:ME2011}%
  \BibitemOpen
  \bibfield  {author} {\bibinfo {author} {\bibfnamefont {G.}~\bibnamefont
  {Niu}}, \bibinfo {author} {\bibfnamefont {S.}~\bibnamefont {Yin}}, \bibinfo
  {author} {\bibfnamefont {G.}~\bibnamefont {Saint-Girons}}, \bibinfo {author}
  {\bibfnamefont {B.}~\bibnamefont {Gautier}}, \bibinfo {author} {\bibfnamefont
  {P.}~\bibnamefont {Lecoeur}}, \bibinfo {author} {\bibfnamefont
  {V.}~\bibnamefont {Pillard}}, \bibinfo {author} {\bibfnamefont
  {G.}~\bibnamefont {Hollinger}}, \ and\ \bibinfo {author} {\bibfnamefont
  {B.}~\bibnamefont {Vilquin}},\ }\href
  {http://dx.doi.org/10.1016/j.mee.2011.03.028} {\bibfield  {journal} {\bibinfo
   {journal} {Microelectronic Engineering},\ }\textbf {\bibinfo {volume}
  {88}},\ \bibinfo {pages} {1232} (\bibinfo {year} {2011})}\BibitemShut
  {NoStop}%
\bibitem [{\citenamefont {Barbier}\ \emph {et~al.}(2005)\citenamefont
  {Barbier}, \citenamefont {Belkhou}, \citenamefont {Ohresser}, \citenamefont
  {Gautier\-Soyer}, \citenamefont {Bezencenet}, \citenamefont {Mulazzi},
  \citenamefont {Guittet},\ and\ \citenamefont {Moussy}}]{barbier2005}%
  \BibitemOpen
  \bibfield  {author} {\bibinfo {author} {\bibfnamefont {A.}~\bibnamefont
  {Barbier}}, \bibinfo {author} {\bibfnamefont {R.}~\bibnamefont {Belkhou}},
  \bibinfo {author} {\bibfnamefont {P.}~\bibnamefont {Ohresser}}, \bibinfo
  {author} {\bibfnamefont {M.}~\bibnamefont {Gautier\-Soyer}}, \bibinfo
  {author} {\bibfnamefont {O.}~\bibnamefont {Bezencenet}}, \bibinfo {author}
  {\bibfnamefont {M.}~\bibnamefont {Mulazzi}}, \bibinfo {author} {\bibfnamefont
  {M.~J.}\ \bibnamefont {Guittet}}, \ and\ \bibinfo {author} {\bibfnamefont
  {J.~B.}\ \bibnamefont {Moussy}},\ }\href@noop {} {\bibfield  {journal}
  {\bibinfo  {journal} {Phys.\ Rev. B},\ }\textbf {\bibinfo {volume} {72}},\
  \bibinfo {pages} {245423} (\bibinfo {year} {2005})}\BibitemShut {NoStop}%
\bibitem [{\citenamefont {Barbier}\ \emph {et~al.}(2007)\citenamefont
  {Barbier}, \citenamefont {Bezencenet}, \citenamefont {Mocuta}, \citenamefont
  {Moussy}, \citenamefont {Magnan}, \citenamefont {Jedrecy}, \citenamefont
  {Guittet},\ and\ \citenamefont {Gautier-Soyer}}]{barbier2007}%
  \BibitemOpen
  \bibfield  {author} {\bibinfo {author} {\bibfnamefont {A.}~\bibnamefont
  {Barbier}}, \bibinfo {author} {\bibfnamefont {O.}~\bibnamefont {Bezencenet}},
  \bibinfo {author} {\bibfnamefont {C.}~\bibnamefont {Mocuta}}, \bibinfo
  {author} {\bibfnamefont {J.~B.}\ \bibnamefont {Moussy}}, \bibinfo {author}
  {\bibfnamefont {H.}~\bibnamefont {Magnan}}, \bibinfo {author} {\bibfnamefont
  {N.}~\bibnamefont {Jedrecy}}, \bibinfo {author} {\bibfnamefont {M.~J.}\
  \bibnamefont {Guittet}}, \ and\ \bibinfo {author} {\bibfnamefont
  {M.}~\bibnamefont {Gautier-Soyer}},\ }\href@noop {} {\bibfield  {journal}
  {\bibinfo  {journal} {Mat. Sci and Eng.},\ }\textbf {\bibinfo {volume} {B
  144}},\ \bibinfo {pages} {19} (\bibinfo {year} {2007})}\BibitemShut {NoStop}%
\bibitem [{DIF()}]{DIFFABS}%
  \BibitemOpen
  \href@noop {} {}\bibinfo {howpublished}
  {\url{http://www.synchrotron-soleil.fr/portal/page/portal/Recherche/LignesLu%
miere/DIFFABS}}\BibitemShut {NoStop}%
\bibitem [{\citenamefont {de~Ligny}\ and\ \citenamefont
  {Richet}(1996)}]{Ligny:Phys.Rev.B.1995}%
  \BibitemOpen
  \bibfield  {author} {\bibinfo {author} {\bibfnamefont {D.}~\bibnamefont
  {de~Ligny}}\ and\ \bibinfo {author} {\bibfnamefont {P.}~\bibnamefont
  {Richet}},\ }\href@noop {} {\bibfield  {journal} {\bibinfo  {journal} {Phys.
  Rev. B},\ }\textbf {\bibinfo {volume} {53}},\ \bibinfo {pages} {3013}
  (\bibinfo {year} {1996})}\BibitemShut {NoStop}%
\bibitem [{\citenamefont {Gorev}\ \emph {et~al.}(2009)\citenamefont {Gorev},
  \citenamefont {Bondarev}, \citenamefont {Flerov}, \citenamefont {Maglione},
  \citenamefont {Simon}, \citenamefont {Sciau}, \citenamefont {Boulos},\ and\
  \citenamefont {Guillemet-Fritsch}}]{Gorev:Phys.Rev.B.1995}%
  \BibitemOpen
  \bibfield  {author} {\bibinfo {author} {\bibfnamefont {M.}~\bibnamefont
  {Gorev}}, \bibinfo {author} {\bibfnamefont {V.}~\bibnamefont {Bondarev}},
  \bibinfo {author} {\bibfnamefont {I.}~\bibnamefont {Flerov}}, \bibinfo
  {author} {\bibfnamefont {M.}~\bibnamefont {Maglione}}, \bibinfo {author}
  {\bibfnamefont {A.}~\bibnamefont {Simon}}, \bibinfo {author} {\bibfnamefont
  {P.}~\bibnamefont {Sciau}}, \bibinfo {author} {\bibfnamefont
  {M.}~\bibnamefont {Boulos}}, \ and\ \bibinfo {author} {\bibfnamefont
  {S.}~\bibnamefont {Guillemet-Fritsch}},\ }\Doi
  {10.1088/0953-8984/21/7/075902} {\bibfield  {journal} {\bibinfo  {journal}
  {J. Phys.: Condens. Matter},\ }\textbf {\bibinfo {volume} {21}},\ \bibinfo
  {pages} {075902} (\bibinfo {year} {2009})}\BibitemShut {NoStop}%
\bibitem [{\citenamefont {Rao}\ and\ \citenamefont
  {Umarji}(1997)}]{Rao:BMS.1997}%
  \BibitemOpen
  \bibfield  {author} {\bibinfo {author} {\bibfnamefont {M.~V.~R.}\
  \bibnamefont {Rao}}\ and\ \bibinfo {author} {\bibfnamefont {A.}~\bibnamefont
  {Umarji}},\ }\href@noop {} {\bibfield  {journal} {\bibinfo  {journal} {Bull.
  Mater. Sci.},\ }\textbf {\bibinfo {volume} {20}},\ \bibinfo {pages} {1023}
  (\bibinfo {year} {1997})}\BibitemShut {NoStop}%
\bibitem [{\citenamefont {Schubert}\ \emph {et~al.}(2003)\citenamefont
  {Schubert}, \citenamefont {Trithaveesak}, \citenamefont {Petraru},
  \citenamefont {Jia}, \citenamefont {Uecker}, \citenamefont {Reiche},\ and\
  \citenamefont {Schlom}}]{Schubert:Appl.Phys.Lett.2003}%
  \BibitemOpen
  \bibfield  {author} {\bibinfo {author} {\bibfnamefont {J.}~\bibnamefont
  {Schubert}}, \bibinfo {author} {\bibfnamefont {O.}~\bibnamefont
  {Trithaveesak}}, \bibinfo {author} {\bibfnamefont {A.}~\bibnamefont
  {Petraru}}, \bibinfo {author} {\bibfnamefont {C.~L.}\ \bibnamefont {Jia}},
  \bibinfo {author} {\bibfnamefont {R.}~\bibnamefont {Uecker}}, \bibinfo
  {author} {\bibfnamefont {P.}~\bibnamefont {Reiche}}, \ and\ \bibinfo {author}
  {\bibfnamefont {D.~G.}\ \bibnamefont {Schlom}},\ }\href@noop {} {\bibfield
  {journal} {\bibinfo  {journal} {Appl. Phys. Lett.},\ }\textbf {\bibinfo
  {volume} {82}},\ \bibinfo {pages} {3460} (\bibinfo {year}
  {2003})}\BibitemShut {NoStop}%
\bibitem [{\citenamefont {Siegert}\ \emph {et~al.}(2000)\citenamefont
  {Siegert}, \citenamefont {Lisoni}, \citenamefont {Lei}, \citenamefont
  {Eckau}, \citenamefont {Zander}, \citenamefont {Jia}, \citenamefont
  {J.Schubert},\ and\ \citenamefont {Buchal}}]{Siegert:MRS.2000}%
  \BibitemOpen
  \bibfield  {author} {\bibinfo {author} {\bibfnamefont {M.}~\bibnamefont
  {Siegert}}, \bibinfo {author} {\bibfnamefont {J.~G.}\ \bibnamefont {Lisoni}},
  \bibinfo {author} {\bibfnamefont {C.~H.}\ \bibnamefont {Lei}}, \bibinfo
  {author} {\bibfnamefont {A.}~\bibnamefont {Eckau}}, \bibinfo {author}
  {\bibfnamefont {W.}~\bibnamefont {Zander}}, \bibinfo {author} {\bibfnamefont
  {C.~L.}\ \bibnamefont {Jia}}, \bibinfo {author} {\bibnamefont {J.Schubert}},
  \ and\ \bibinfo {author} {\bibfnamefont {C.}~\bibnamefont {Buchal}},\
  }\href@noop {} {\bibfield  {journal} {\bibinfo  {journal} {Mater. Res. Soc.
  Symp. Proc.},\ }\textbf {\bibinfo {volume} {597}},\ \bibinfo {pages} {145}
  (\bibinfo {year} {2000})}\BibitemShut {NoStop}%
\bibitem [{\citenamefont {Sun}\ \emph {et~al.}(2004)\citenamefont {Sun},
  \citenamefont {Tian}, \citenamefont {Pan}, \citenamefont {Haeni},\ and\
  \citenamefont {Schlom}}]{Sun:APL2004}%
  \BibitemOpen
  \bibfield  {author} {\bibinfo {author} {\bibfnamefont {H.~P.}\ \bibnamefont
  {Sun}}, \bibinfo {author} {\bibfnamefont {W.}~\bibnamefont {Tian}}, \bibinfo
  {author} {\bibfnamefont {X.~Q.}\ \bibnamefont {Pan}}, \bibinfo {author}
  {\bibfnamefont {J.~H.}\ \bibnamefont {Haeni}}, \ and\ \bibinfo {author}
  {\bibfnamefont {D.~G.}\ \bibnamefont {Schlom}},\ }\Doi {10.1063/1.1728300}
  {\bibfield  {journal} {\bibinfo  {journal} {Appl. Phys. Lett.},\ }\textbf
  {\bibinfo {volume} {84}},\ \bibinfo {pages} {3298} (\bibinfo {year}
  {2004})}\BibitemShut {NoStop}%
\bibitem [{\citenamefont {Shimoyama}\ \emph {et~al.}(2001)\citenamefont
  {Shimoyama}, \citenamefont {Kubo}, \citenamefont {Maeda},\ and\ \citenamefont
  {Yamabe}}]{Shimoyama:2001}%
  \BibitemOpen
  \bibfield  {author} {\bibinfo {author} {\bibfnamefont {K.}~\bibnamefont
  {Shimoyama}}, \bibinfo {author} {\bibfnamefont {K.}~\bibnamefont {Kubo}},
  \bibinfo {author} {\bibfnamefont {T.}~\bibnamefont {Maeda}}, \ and\ \bibinfo
  {author} {\bibfnamefont {K.}~\bibnamefont {Yamabe}},\ }\href@noop {}
  {\bibfield  {journal} {\bibinfo  {journal} {Jpn. J. Appl. Phys.},\ }\textbf
  {\bibinfo {volume} {40}},\ \bibinfo {pages} {L463} (\bibinfo {year}
  {2001})}\BibitemShut {NoStop}%
\bibitem [{\citenamefont {Mukhopadhyay}\ and\ \citenamefont
  {Chen}(1995)}]{Mukhopadhyay:1995}%
  \BibitemOpen
  \bibfield  {author} {\bibinfo {author} {\bibfnamefont {S.~M.}\ \bibnamefont
  {Mukhopadhyay}}\ and\ \bibinfo {author} {\bibfnamefont {C.~S.}\ \bibnamefont
  {Chen}},\ }\href@noop {} {\bibfield  {journal} {\bibinfo  {journal} {J.
  Mater. res.},\ }\textbf {\bibinfo {volume} {10}},\ \bibinfo {pages} {1502}
  (\bibinfo {year} {1995})}\BibitemShut {NoStop}%
\bibitem [{\citenamefont {Henrich}\ \emph {et~al.}(1978)\citenamefont
  {Henrich}, \citenamefont {Dresselhaus},\ and\ \citenamefont
  {Zeiger}}]{Henrich:1978}%
  \BibitemOpen
  \bibfield  {author} {\bibinfo {author} {\bibfnamefont {V.~E.}\ \bibnamefont
  {Henrich}}, \bibinfo {author} {\bibfnamefont {G.}~\bibnamefont
  {Dresselhaus}}, \ and\ \bibinfo {author} {\bibfnamefont {H.~J.}\ \bibnamefont
  {Zeiger}},\ }\href {http://link.aps.org/doi/10.1103/PhysRevB.17.4908}
  {\bibfield  {journal} {\bibinfo  {journal} {Phys. Rev. B},\ }\textbf
  {\bibinfo {volume} {17}},\ \bibinfo {pages} {4908–4921} (\bibinfo {year}
  {1978})}\BibitemShut {NoStop}%
\bibitem [{\citenamefont {Solomon}\ and\ \citenamefont
  {Baum}(1975)}]{Solomon:1975}%
  \BibitemOpen
  \bibfield  {author} {\bibinfo {author} {\bibfnamefont {J.~S.}\ \bibnamefont
  {Solomon}}\ and\ \bibinfo {author} {\bibfnamefont {W.~L.}\ \bibnamefont
  {Baum}},\ }\href {http://dx.doi.org/10.1016/0039-6028(75)90245-9} {\bibfield
  {journal} {\bibinfo  {journal} {Surf. Sci.},\ }\textbf {\bibinfo {volume}
  {51}},\ \bibinfo {pages} {228} (\bibinfo {year} {1975})}\BibitemShut
  {NoStop}%
\bibitem [{\citenamefont {Verhoeven}(1979)}]{Verhoeven:1979}%
  \BibitemOpen
  \bibfield  {author} {\bibinfo {author} {\bibfnamefont {J.}~\bibnamefont
  {Verhoeven}},\ }\href {http://dx.doi.org/10.1016/S0042-207X(80)80046-7}
  {\bibfield  {journal} {\bibinfo  {journal} {Vacuum},\ }\textbf {\bibinfo
  {volume} {30}},\ \bibinfo {pages} {69} (\bibinfo {year} {1979})}\BibitemShut
  {NoStop}%
\bibitem [{Sim()}]{Simul}%
  \BibitemOpen
  \href@noop {} {}\bibinfo {howpublished}
  {\url{http://www-llb.cea.fr/prism/programs/simulreflec
  /simulreflec.html}}\BibitemShut {NoStop}%
\bibitem [{\citenamefont {Nagarajan}\ \emph {et~al.}(2004)\citenamefont
  {Nagarajan}, \citenamefont {Tian}, \citenamefont {Pan}, \citenamefont {Kim},
  \citenamefont {Eom}, \citenamefont {Kohlstedt},\ and\ \citenamefont
  {Waser}}]{Nagarajan:2004}%
  \BibitemOpen
  \bibfield  {author} {\bibinfo {author} {\bibfnamefont {V.}~\bibnamefont
  {Nagarajan}}, \bibinfo {author} {\bibfnamefont {W.}~\bibnamefont {Tian}},
  \bibinfo {author} {\bibfnamefont {X.~Q.}\ \bibnamefont {Pan}}, \bibinfo
  {author} {\bibfnamefont {D.~M.}\ \bibnamefont {Kim}}, \bibinfo {author}
  {\bibfnamefont {C.~B.}\ \bibnamefont {Eom}}, \bibinfo {author} {\bibfnamefont
  {H.}~\bibnamefont {Kohlstedt}}, \ and\ \bibinfo {author} {\bibfnamefont
  {R.}~\bibnamefont {Waser}},\ }\href@noop {} {\bibfield  {journal} {\bibinfo
  {journal} {Appl. Phys. Lett.},\ }\textbf {\bibinfo {volume} {84}},\ \bibinfo
  {pages} {5225} (\bibinfo {year} {2004})}\BibitemShut {NoStop}%
\bibitem [{\citenamefont {Allibe}\ \emph {et~al.}(2012)\citenamefont {Allibe},
  \citenamefont {Fusil}, \citenamefont {Bouzehouane}, \citenamefont {Daumont},
  \citenamefont {Sando}, \citenamefont {Jacquet}, \citenamefont {Deranlot},
  \citenamefont {Bibes},\ and\ \citenamefont
  {A.Barth\'{e}l\`{e}my}}]{Allibe:2012}%
  \BibitemOpen
  \bibfield  {author} {\bibinfo {author} {\bibfnamefont {J.}~\bibnamefont
  {Allibe}}, \bibinfo {author} {\bibfnamefont {S.}~\bibnamefont {Fusil}},
  \bibinfo {author} {\bibfnamefont {K.}~\bibnamefont {Bouzehouane}}, \bibinfo
  {author} {\bibfnamefont {C.}~\bibnamefont {Daumont}}, \bibinfo {author}
  {\bibfnamefont {D.}~\bibnamefont {Sando}}, \bibinfo {author} {\bibfnamefont
  {E.}~\bibnamefont {Jacquet}}, \bibinfo {author} {\bibfnamefont
  {C.}~\bibnamefont {Deranlot}}, \bibinfo {author} {\bibfnamefont
  {M.}~\bibnamefont {Bibes}}, \ and\ \bibinfo {author} {\bibnamefont
  {A.Barth\'{e}l\`{e}my}},\ }\href@noop {} {\bibfield  {journal} {\bibinfo
  {journal} {NanoLett.},\ }\textbf {\bibinfo {volume} {12}},\ \bibinfo {pages}
  {1141} (\bibinfo {year} {2012})}\BibitemShut {NoStop}%
\bibitem [{\citenamefont {Davis}\ \emph {et~al.}(1977)\citenamefont {Davis},
  \citenamefont {MacDonald}, \citenamefont {Palmberg}, \citenamefont {Riach},\
  and\ \citenamefont {R.E.Weber}}]{AES_Handbook}%
  \BibitemOpen
  \bibinfo {editor} {\bibfnamefont {L.}~\bibnamefont {Davis}}, \bibinfo
  {editor} {\bibfnamefont {N.}~\bibnamefont {MacDonald}}, \bibinfo {editor}
  {\bibfnamefont {P.}~\bibnamefont {Palmberg}}, \bibinfo {editor}
  {\bibfnamefont {G.}~\bibnamefont {Riach}}, \ and\ \bibinfo {editor}
  {\bibnamefont {R.E.Weber}},\ eds.,\ \href@noop {} {\emph {\bibinfo {title}
  {Handbook of Auger Electron Spectroscopy (second edition)}}}\ (\bibinfo
  {publisher} {Physical Electronics Division, Perkin Elmer Corporation, Eden
  Prairie, Minnesota},\ \bibinfo {year} {1977})\BibitemShut {NoStop}%
\bibitem [{\citenamefont {Li}\ \emph {et~al.}(2008)\citenamefont {Li},
  \citenamefont {Lu}, \citenamefont {Li}, \citenamefont {Mai}, \citenamefont
  {Kim},\ and\ \citenamefont {Jia}}]{Li:APL2008}%
  \BibitemOpen
  \bibfield  {author} {\bibinfo {author} {\bibfnamefont {X.~L.}\ \bibnamefont
  {Li}}, \bibinfo {author} {\bibfnamefont {H.~B.}\ \bibnamefont {Lu}}, \bibinfo
  {author} {\bibfnamefont {M.}~\bibnamefont {Li}}, \bibinfo {author}
  {\bibfnamefont {Z.}~\bibnamefont {Mai}}, \bibinfo {author} {\bibfnamefont
  {H.}~\bibnamefont {Kim}}, \ and\ \bibinfo {author} {\bibfnamefont {Q.~J.}\
  \bibnamefont {Jia}},\ }\href@noop {} {\bibfield  {journal} {\bibinfo
  {journal} {Appl. Phys. Lett.},\ }\textbf {\bibinfo {volume} {92}} (\bibinfo
  {year} {2008})}\BibitemShut {NoStop}%
\end{thebibliography}
\end{document}